# Demagnetization Protocols for Frustrated Interacting Nanomagnet Arrays

R. F. Wang[1], J. Li[1], W. McConville[1], C. Nisoli[1], X. Ke[1], J.W. Freeland[2], V. Rose[2,3], M. Grimsditch[4], P. Lammert[1], V. H. Crespi[1], and P. Schiffer[1]

[1]Department of Physics and Materials Research Institute, Pennsylvania State University, University Park, PA, 16802

[2]Advanced Photon Source, Argonne National Laboratory, Argonne, IL 60439

[3]Center for Nanoscale Materials, Argonne National Laboratory, Argonne, IL 60439

[4]Materials Science Division, Argonne National Laboratory, Argonne, IL 60439

## ABSTRACT

We report a study of demagnetization protocols for frustrated arrays of interacting single domain permalloy nanomagnets by rotating the arrays in a changing magnetic field. The most effective demagnetization is achieved by not only stepping the field strength down while the sample is rotating, but by combining each field step with an alternation in the field direction. By contrast, linearly decreasing the field strength or stepping the field down without alternating the field direction leaves the arrays with a larger remanent magnetic moment. These results suggest that non-monotonic variations in field magnitude around and below the coercive field are important for the demagnetization process.

Nanometer sized ferromagnets are both of fundamental interest[1,2] and importance to a variety of technological applications. [3,4,5,6] Previous studies showed that arrays or chains of nanomagnets [7, 8, 9, 10] displayed various magnetization patterns after they were demagnetized and these patterns were closely related to the nature of the magnetic interactions between neighboring nanomagnets. We have studied demagnetization protocols for geometrically frustrated square arrays of single-domain permalloy nanomagnets.[11] We find that rotation in a magnetic field which also alternates in direction while its amplitude is decreased can effectively demagnetize the arrays, but that rotation in a magnetic field with continuously diminishing amplitude is less effective.

We fabricated the frustrated arrays by electron beam lithography, as shown in figure 1 and described in detail elsewhere.[11] The permalloy islands were approximately 25 nm thick, 80 nm wide, and 220 nm long, with a range of different lattice constants (from 320 to 880 nm) for the square array. The single-domain nature of the permalloy magnets is revealed by simple black-white contrast in the MFM image shown in figure 1b. The moments of these nanomagnets aligned with the islands' long axes due to the strong shape anisotropy. Figure 2 displays the hysteresis loops of the four arrays obtained by magneto-optical Kerr effect (MOKE)[12] measurements. The arrays had nearly the same coercive field, about 770 Oe, regardless of their lattice constant.

During demagnetization, our sample was placed on a 1000 rpm rotating stage inside a changing magnetic field, an approach similar to the one used in reference 10. The demagnetization procedure always began at a field of 1,300 Oe or higher (well above the easy-axis coercive field, as indicated by the open part of the hysteresis loop of figure 2). We tested three different demagnetization protocols, as shown in figure 3 (top panel). In protocol 1 (used in reference 11) [13], the magnetic field strength was stepped down in magnitude and switched polarity with each step. In protocols 2 and 3, the magnetic field was decreased to zero either linearly (protocol 2) or by steps (protocol 3) without changing polarity. We could adjust independently the field ramping rate ($R_m$) the field step size ($S_H$), and the time that the array stayed at each field ($T_s$).

After each demagnetization procedure, we took magnetic force microscope (MFM) images of the square lattices in zero magnetic field. Each MFM image covered an area of 10 by 10 microns (including about 280 to 1300 permalloy islands in each MFM image, depending on lattice constant). Because the moment of each island is parallel to its long axis, we assigned a value of ±1 to each island moment (positive being defined as upward or to the right, depending on the island orientation). After mapping the moments of all islands in the MFM image, we could count the numbers of upward and downward moments ($N_y$ and $N_{-y}$) as well as rightward and leftward moments ($N_x$ and $N_{-x}$). For each MFM image, we calculated the remanent magnetization as:

$$m_y = (N_y - N_{-y}) / (N_y + N_{-y})$$

$$m_x = (N_x - N_{-x}) / (N_x + N_{-x})$$

Total remanent magnetization: $\quad m_{tot} = \sqrt{m_x^2 + m_y^2} / \sqrt{2}$

Under this definition, $m_{tot} = 1$ would be a fully magnetized state, and we use $m_{tot}$ to characterize the efficiency of the demagnetization protocols.

In protocol 1, we set the magnetic field sequence as $H_1, -H_2, H_3, -H_4, ..., 0$ where the negative field values had switched polarity in the laboratory frame. We defined the field step as $S_H = |H_i| - |H_{i+1}|$ in this case. We used $R_m = 24{,}000$ Oe/second and $T_s = 1$ s for most of these tests (a small number of tests with slower ramp rate and longer hold times suggest that these factors do not strongly affect the demagnetization). In this protocol we used $S_H$ ~ 32.6 Oe (for $H$ = 1308 Oe to -816 Oe ), followed by steps to 800 Oe, -767 Oe, 734 Oe, and then $S_H$ ~ 16.3 Oe down to $H$ = 0. We used protocol 1 to demagnetize our samples multiple times for each lattice spacing, measuring the orientation of over 2000 islands for each array lattice spacing except the largest (880 nm) where we imaged only 1100 due to the smaller number in each MFM image.

Repeated runs on an array with lattice parameter of 320 nm revealed that 49.2% of 3568 islands kept their initial orientation after repeating the protocol (the corresponding number was 49.3% in an array with lattice parameter of 560 nm). This result is very close to the ideal value of 50%, and it strongly suggests that there is no sample history dependence to the results of the demagnetization. As shown in figure 3 (bottom panel), $m_{tot}$ ranges between 0.056 and 0.152 for the 8 lattices, indicating rather good demagnetization, although there is an apparent slight increase in $m_{tot}$ as the lattice constant increases. This apparent increase may simply be a statistical effect associated with the 10 x 10 micron size of all of the MFM images. This fixed size results in about 1300 permalloy islands in one MFM image for the 320 nm array, compared to only 280 in the 880 nm array. Alternatively, the weak trend may be associated with the larger island-island interaction in arrays of the smaller lattice constant. The left panel of Figure 4 displays an array with a lattice constant of 360 nm after demagnetization by protocol 1. The inset to that panel shows a two-dimensional discrete Fourier transform of the same image with each magnetic moment represented by ±1. The small remanent magnetization for protocol 1 is seen in the weak intensity in the center of the Fourier transformed image.

In contrast to the effective and reproducible demagnetization through protocol 1, protocols 2 and 3 were not as effective at demagnetizing the arrays for any lattice constant. In the linear ramp of protocol 2, we tested $R_m$ = 8, 0.32, and 0.08 Oe/second, which typically produced $m_{tot}$ of at least twice that produced by protocol 1 when tested on the same arrays, and some instances these protocols produced magnetizations as high as $m_{tot}$ ~ 0.5 (which is still well below the full saturation magnetization of $m_{tot}$ = 1). The right panel of figure 4 shows the MFM image of an array with lattice constant of 360 nm after demagnetization by protocol 2; note the large central peak in the Fourier transform. Protocol 3, with the same step sizes as protocol 1, resulted in a similarly large magnetization range as protocol 2.

The results of our investigation indicate that nonmonotonic excursions in the magnetic field strength within a decreasing field envelope substantially improve demagnetization of the arrays for protocols of the step sizes studied. This suggests that the process of magnetizing the arrays, taking the field below the coercive field, and then remagnetizing with a field near the coercive field is the important factor in optimizing demagnetization. The efficacy of sweeping through zero field further suggests that there is a broad range of time and length scales which are of importance to the demagnetization process. This would be expected in a glassy system, as our frustrated arrays may be, but a more detailed local investigation of the collective dynamics or studies of well-isolated islands could shed considerable additional light on whether this is relevant here. The present results do, however, demonstrate the need to follow a careful protocol in the demagnetization of arrays of single-domain nanomagnets – a necessary step in order to study the important many-body effects resulting from magnetostatic interactions within such arrays.

We acknowledge the financial support from Army Research Office and the National Science Foundation MRSEC program. Work at Argonne was supported by the U.S. Department of Energy, Office of Science, Office of Basic Energy Sciences, under Contract No. W-31-109-Eng-38. We are grateful for sample preparation assistance from M. S. Lund and C. Leighton.

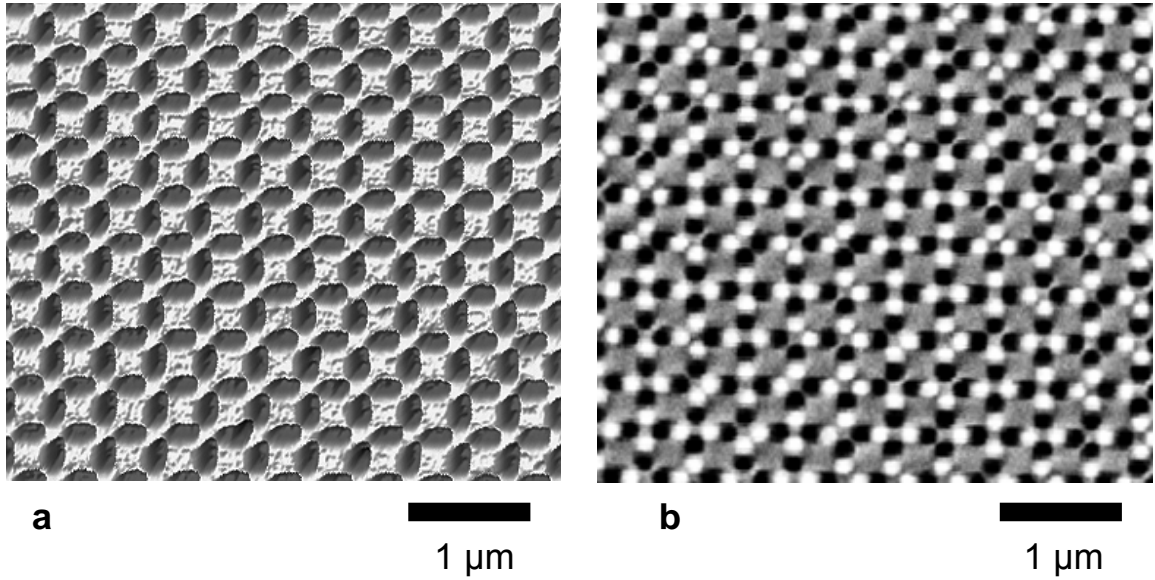

Figure 1. 1a). AFM image of a square array of permalloy nanomagnets with lattice constant of 400nm. 1b). Corresponding MFM image covering the same area of the array after demagnetization with protocol 1. All permalloy islands are in the single domain state. The magnetization direction points from white to black.

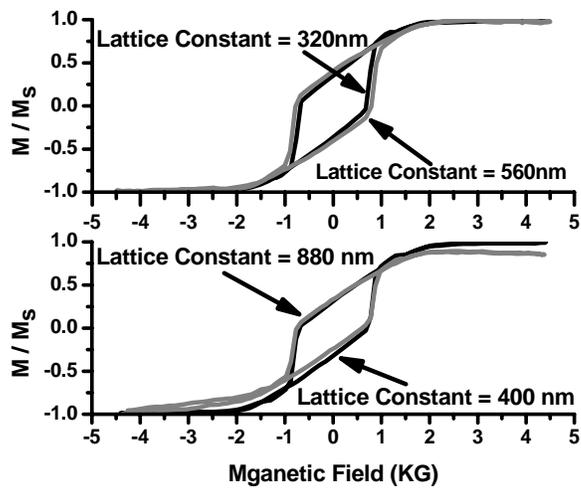

Figure 2. Hysteresis loop, measured by MOKE, of square arrays with lattice constant of 320nm, 560nm (upper panel), 400nm, and 880nm (lower panel).

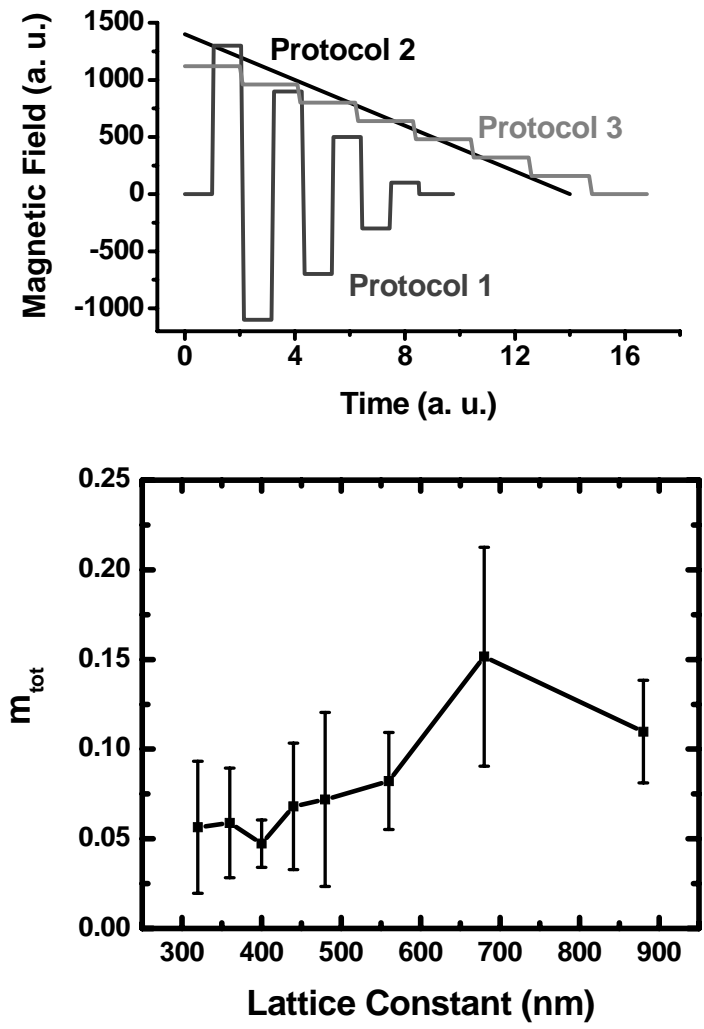

**Figure 3. top:** Schematic plots of our demagnetization protocols. **bottom:** The total magnetization of the arrays as a function of the array lattice constant, averaging several MFM images. The error bars are the standard deviation of the average of the images.

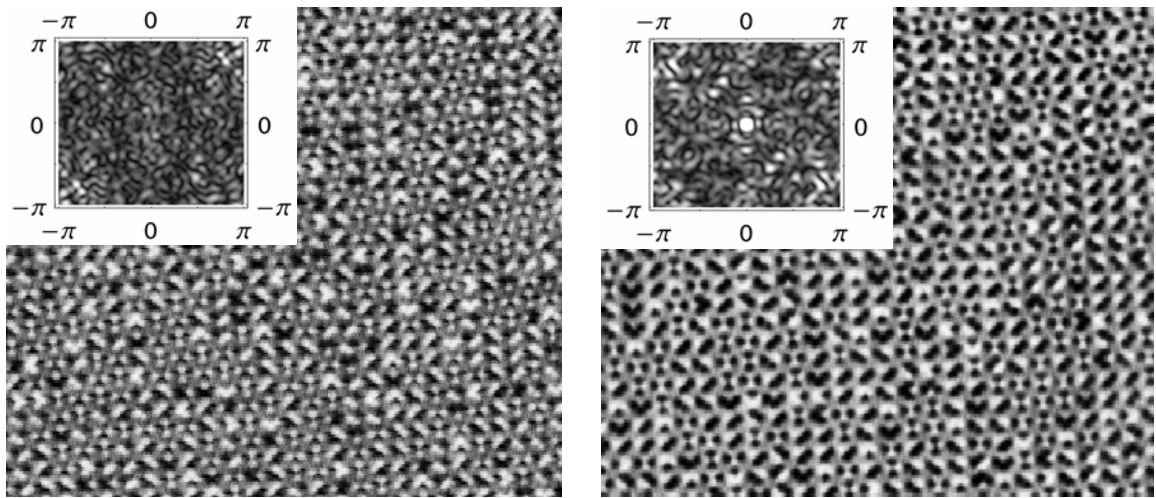

Figure 4. MFM images and their Fourier transformed images of a square array with lattice constant of 360nm after demagnetization, using protocol 1 (left panel), protocol 2 (right panel).

## References

[1] J. I. Martin, J. Nogues, Kai Liu, J. L. Vicent, and Ivan K. Schuller, J. Magn. Magn. Mater. **256** 449 (2003) and the references therein.

[2] S. D. Bader, Rev. Mod. Phys. **78**, 6861 (2006) and the references therein.

[3] Stephen Y. Chou, Peter R. Krauss, and Linshu Kong, J. Appl. Phys. **79**, 6101 (1996).

[4] C. A. Ross, Henry I. Smith, T. Savas, M. Schattenburg, M. Farhoud, M. Hwang, M. Walsh, M. C. Abraham, and R. J. Ram, J. Vac. Sci. Technol. B **17**, 3168 (1999).

[5] Jian-Gang Zhu, Youfeng Zheng, and Gary A. Prinz, J. Appl. Phys. **87**, 6668 (2000)

[6] R. P. Cowburn and M. E. Welland, Science **287**, 1466 (2000).

[7] M. Hwang, M. C. Abraham, T. A. Savas, Henry I. Smith, R. J. Ram, and C. A. Ross, J. Appl. Phys. **87**, 5108 (2000).

[8] T. Aign, P. Meyer, S. Lemerle, J. P. Jamet, J. Ferre, V. Mathet, C. Chappert, J. Gierak, C. Vieu, F. Rousseaus, H. Launois, and H. Bernas, Phys. Rev. Lett. **81**, 5656 (1998).

[9] R. P. Cowburn, Phys. Rev. B **65**, 092409 (2002).

[10] Alexandra Imre, Gyorgy Csaba, Gary H. Bernstein, Wolfgang Porod, and Vitali Metlushko, Superlatt Microstruct **34**, 513 (2004).

[11] R. F. Wang, C. Nisoli, R. S. Freitas, J. Li, W. McConville, B. J. Cooley, M. S. Lund, N. Samarth, C. Leighton, V. H. Crespi, and P. Schiffer, Nature **439**, 303 (2006).

[12] M. Grimsditch, and P. Vavassori, J. Phys. : Condens. Matter **16**, R275 (2004)

[13] At the time of publication of reference 11, the importance of switching the polarity of the magnetic field in the laboratory reference frame was not recognized. As a result, due to the complex change in direction of the magnetic field relative to the sample during

protocol 1, it was simply described as the magnetic field "…gradually stepping down in magnitude to zero" in that publication.